# Is Power Law Scaling a quantitative description of Darwin Theory of Evolution?

**Hari Mohan Gupta**
**José Roberto Campanha**
**Dpto Física – UNESP - Rio Claro – São Paulo**
**CCComplex – Virtual Research Center**
**Science and Computing for Complexity**

## Abstract

In the present work, via computational simulation we study the statistical distribution of people versus number of steps acquired by them in a learning process, considering Darwin classical theory of evolution, i.e. competition, learning and survival for the fittest. We consider that learning ability is normally distributed. We found that the number of people versus step acquired by them in a learning process is given through a power law ($N(n) \approx cn^{-\alpha}$). As competition, learning and survival for the fittest is also at the heart of all economical and social systems, we consider that in some cases, power law scaling is a quantitative description of Darwin theory of evolution. This gives an alternative thinking in holistic properties of complex systems.

## I. Introduction

Only recently physicists started to study the natural systems as a whole rather than in parts [1-6] and are interested in holistic properties of these systems normally called as "Complex Systems" since they are difficult to understand them from basic principles. The difficulties in understanding these systems arises from the fact that a large number of elementary interactions are taking place at the same time for a large number of components. Further, these systems are in constant evolution and do not have an equilibrium state [1]. Thus the general equilibrium theories like that of perfect markets, perfect rationality etc…, are not applicable. Properties of these systems are much different than the properties of individual sub-units and thus, a holistic description become necessary. Power law scaling [7,8] is observed in many such biological [9-11], physical [2,12-20] and socio-economical systems [21-29] and is now considered as important propertie of these systems.

Physicists by nature believe in explaining everything in terms of basic interactions, which is important in explaining simple laboratory made systems and many technological developments [30]. Various attempts have also been made to explain these complex systems in term of basic interactions. With this objective, the Ising model, Monte Carlo simulations and Mean Field approximation [31-32] has been used in many cases to explain certain properties of these systems. Tsallis [33-35], on the other hand extended Boltzmann-Gibbs statistics to take into account long-range interactions and microscopic memory. He gives a new definition of non-extensive entropy and is able to explain power law scaling.

Bak [1,36] in his theory of "Self Organized Criticality" considered that natural systems self organizes to complex critical state without interference of any outside agent. The process of self organization takes place over a very long transient period. Complex behavior is always created by a long period of evolution. Zipf [37] explained power law as a consequence of human behavior and the principle of least efforts in social systems.

All these complex natural systems have also been discussed in their own field in a semi-quantitative way. In biological sciences, the process of evolution is an area of great importance and Darwin classical theory of evolution is considered as a basis of biological evolution. In this theory, evolution process is discussed considering constant evolution, competition and "survival for the fittest". The interesting point to emphasize is that competition, learning and survival for the fittest are in heart of all economical, historical and social systems too.

In the present paper, we present a model of evolution, i.e. learning and natural selection for the fittest to obtain statistical distribution of people in a field versus various levels of learning and showed that it leads to a power law distribution. We formulated our problem in terms of learning and selection processes in our social and education system to which we are quite familiar.

In section II, we present our model and final expressions for distribution. In section III, we discuss statistical distribution through computational simulations for various parameters. Finally in section IV, we present our conclusions.

## II. The Model and Calculations.

All animals need some mean to acquire food and fulfill basic needs with security. For this purpose, they acquire a certain degree of ability in some skill or techniques through experience and learning. Let us take the case of human beings. We all acquire certain degree of ability in some field depending on our interest, facility to learn a field and market available for that field. This field can be a technical, commercial or some kind of sport or fine art. The learning depends both on schooling and experience. We can go more and deeper in a field depending on our ability to learn that. At every stage of learning, we have a selection process. Only selected people can have more schooling or more experience through higher positions.

Let us consider that people have normal distribution of their ability "**A**" to learn a particular field. By ability we mean that they learn this fraction of subject in first learning process. We consider that mean value and standard deviation of distribution of A is μ and σ. Thus

$$p(A) = \frac{1}{\sqrt{2\pi}\sigma} \exp[-\frac{(A-\mu)^2}{2\sigma^2}] \qquad (1)$$

where p(A) is the probability density of learning ability equal to A. In first stage, a person learn a fraction A of the given subject. Further

$$f_1(A) = F_1(A) = A \tag{2}$$

Here, $f_n(A)$, denotes learning in the $n^{th}$ stage, while $F_n(A)$ denotes average learning up to $n^{th}$ stage. In the second stage, he also needed the knowledge of material given in the first stage. Thus in this stage he will learn only A times the average knowledge he has earlier. In all professional fields, learning is continuous and depends on earlier knowledge. Thus in the second stage, he will learn

$$f_2(A) = AF_1(A) = A.A = A^2 \tag{3}$$

The average knowledge up to second stage is:

$$F_2(A) = \frac{1}{2}(f_2(A) + F_1(A)) \tag{4}$$

Substituting the value of $f_2(A)$, we obtains:

$$F_2(A) = \frac{F_1(A)}{2}(1+A) = \frac{A}{2}(1+A) \tag{5}$$

The average knowledge acquired by him in $3^{th}$ stage in this way will be:

$$F_3(A) = \frac{F_2(A)}{3}(A+(3-1)) = \frac{A}{6}(1+A)(2+A) \tag{6}$$

Continuing this process, the average learning up to $n^{th}$ stage is given by:

$$F_n(A) = \frac{F_{n-1}(A)}{n}(A+(n-1)) \tag{7}$$

Now we consider the selection process. The society or nature will allow only those people to go to $n^{th}$ stage, who have potential to acquire a certain minimum critical fraction of the knowledge say $x_c$, depending on market availability. Clearly society can not give higher education or better experience through higher positions to all people. The value of $x_c$ depends on the field chosen. For example in commercial and technical fields it must be low as most of the people make their earning through these fields, while in sport and fine arts, it must be high.

The people who have sufficient ability to go to the $n^{th}$ stage, but insufficient to go to $(n+1)^{th}$ stage will acquire knowledge up to $n^{th}$ stage. Thus $A_n$ and $A_{n+1}$ are defined through

$$F_n(A_n) = x_c \tag{8}$$
$$F_{n+1}(A_{n+1}) = x_c \tag{9}$$

The number of people in the $n^{th}$ stage, N(n) is thus given by:

$$N(n) = N_0 \int_{A_n}^{A_{n+1}} P(A)dA \qquad (10)$$

where $N_0$ is a renormalization constant given through

$$N_0 \int_{x_c}^{1} P(A)dA = \text{total number of people in the field} \qquad (11)$$

and N(n) gives the distribution of the number of people vs. acquired stages of learning.

## III. Discussion

For the purpose of illustration, we consider mean ability $\mu=0.5$. The value of $\mu$ is not important for the distribution of N(n). Only the difference between $x_c$ and $\mu$, i.e. $(x_c-\mu)$ is relevant.

In Figure 1, we draw number of people N(n) versus number of steps acquired (n) in log-log scale for different values of standard deviation of ability distribution ($\sigma$) varying from 0.1 to 0.3. We choose cut-off value ($x_c$) equal 0.5.

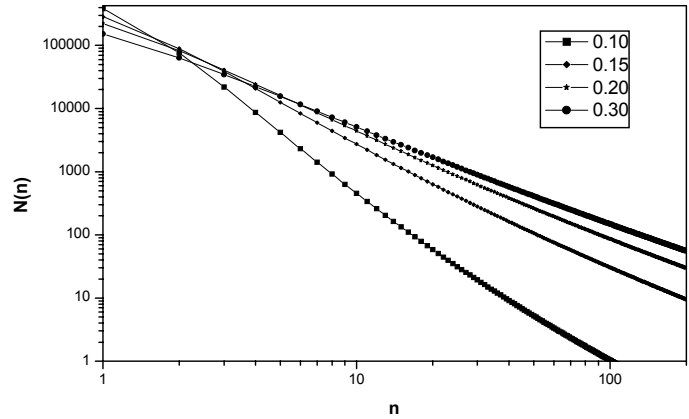

Figure 1 – Variation of number of people N(n) versus number of steps acquired n in log-log scale for different values of standard deviation of ability distribution

In Figure 2 we draw the same for different cut-off values ($x_c$) varying from 0.3 to 0.7, for the standard deviation ($\sigma$) equal to 0.2. We found that the central part in all cases is given through

$$\log N(n) = C - \alpha \log n \qquad (12)$$

which gives a power law distribution.

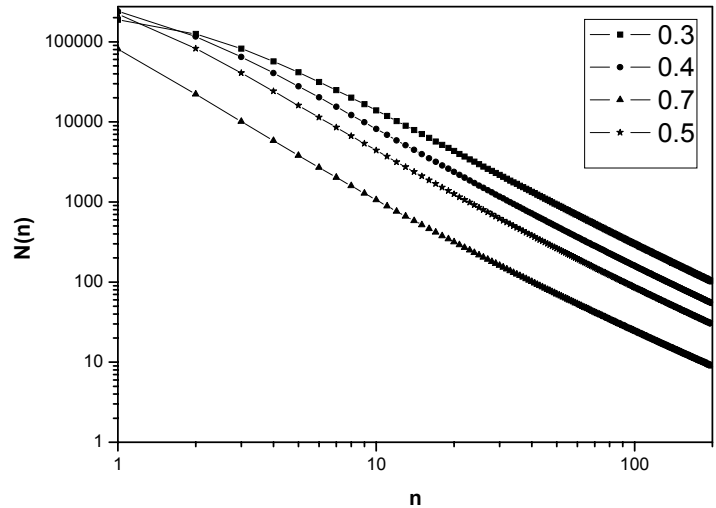

Figure 2 – Variation of number of people N(n) versus number of steps acquired n for different cut-off value $x_c$ in selection process.

For small values of cut-off parameters, i.e., 0.3 and 0.4, the number of people is less than what is given through power law in initial stages. This is also expected by Tsallis statistics and is observed empirically [38].

In Figure 3 we show the variation of power law index α with cut-off value $x_c$ for standard deviation (σ) equal 0.2, while in Figure 4 we shown the variation of power law index α with standard deviation (σ) for the cut-off value ($x_c$) equal to 0.5.

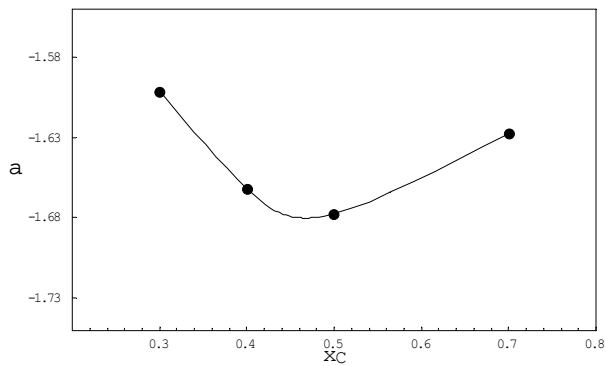

Figure 3 – Variation of power law index β with cut-off value $x_c$.

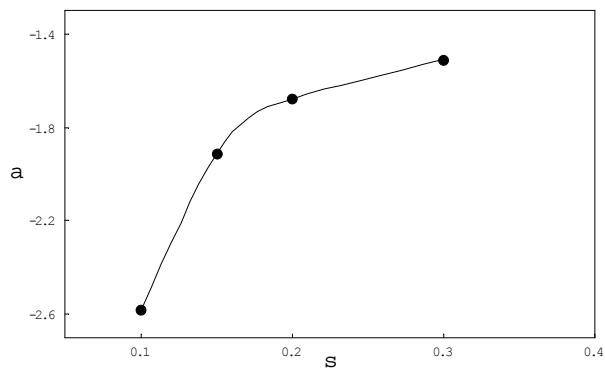

Figure 4 – Variation of power law index β with standard deviation σ.

If we consider that income (or knowledge) of people is proportional to acquired skill, or some power of it, particularly commercial skill, we get power law distribution of income as first obtained by Pareto in 1839.

In the present study, we are considering that all the people has gone to highest stage to which they can go. However the people are evaluating while working particularly for higher stages. The whole system is not in an equilibrium stage. It is constantly evaluating. At any time, many people are in a lower stage than the one they deserve, because they are still in evaluation process. Further many people do not reach to highest stage because of their own desire or some external forces, for example they die or retire. Thus in real world, the number of people in higher stages must be less than what is given by our present analysis. It is necessary to gradually truncate this distribution for higher stages. This explain the less number of people in higher stages than given through a power law distribution [39,40] although in the present analysis, we are getting it more than in a power law.

## IV Conclusions

In the present work, we have shown that although learning ability (including interest) of the people is arbitrary distributed and thus have normal distribution, the learning of the people in the field is having large variation or steps in which the people are distributed according to power law due to continuous learning (or evolution) and selection process at each stage. This gives great variety in each segment of life such as biology, economics, history and culture as observed. To obtain this we used a theory which is well established in a semi-quantitative manner in biology, i.e., competition, learning and natural selection for survival for the fittest. The present approach gives an alternative thinking towards mechanisms behind the power law scaling in nature and a universal law in complex system physics.